\documentclass[%
 reprint,
 amsmath,amssymb,
 aps,
]{revtex4-2}

\usepackage{graphicx}
\usepackage{dcolumn}
\usepackage{bm}
\usepackage{xcolor}

\begin{document}

\preprint{APS/123-QED}

\title{Mode-resolved thermometry of trapped ion with Deep Learning}

\author{Yi Tao}
\thanks{These authors contributed equally to this work.}
 \affiliation{Institute for Quantum Science and Technology, College of Science, National University of Defense Technology, Changsha Hunan 410073, China}
 \affiliation{Interdisciplinary Center for Quantum Information, National University of Defense Technology, Changsha Hunan 410073, China} 
 \affiliation{Hunan Key Laboratory of Mechanism and Technology of Quantum Information, Changsha Hunan 410073, China} 
\author{Ting Chen}%
\thanks{These authors contributed equally to this work.}
 \affiliation{Institute for Quantum Science and Technology, College of Science, National University of Defense Technology, Changsha Hunan 410073, China}
 \affiliation{Interdisciplinary Center for Quantum Information, National University of Defense Technology, Changsha Hunan 410073, China} 
 \affiliation{Hunan Key Laboratory of Mechanism and Technology of Quantum Information, Changsha Hunan 410073, China} 
 \author{Yi Xie}%
 \affiliation{Institute for Quantum Science and Technology, College of Science, National University of Defense Technology, Changsha Hunan 410073, China}
\affiliation{Interdisciplinary Center for Quantum Information, National University of Defense Technology, Changsha Hunan 410073, China} 
\affiliation{Hunan Key Laboratory of Mechanism and Technology of Quantum Information, Changsha Hunan 410073, China} 
 \author{Hongyang Wang}%
 \affiliation{Institute for Quantum Science and Technology, College of Science, National University of Defense Technology, Changsha Hunan 410073, China}
\affiliation{Interdisciplinary Center for Quantum Information, National University of Defense Technology, Changsha Hunan 410073, China} 
\affiliation{Hunan Key Laboratory of Mechanism and Technology of Quantum Information, Changsha Hunan 410073, China} 
 \author{Jie Zhang}%
 \affiliation{Institute for Quantum Science and Technology, College of Science, National University of Defense Technology, Changsha Hunan 410073, China}
\affiliation{Interdisciplinary Center for Quantum Information, National University of Defense Technology, Changsha Hunan 410073, China} 
\affiliation{Hunan Key Laboratory of Mechanism and Technology of Quantum Information, Changsha Hunan 410073, China} 
  \author{Ting Zhang}%
  \affiliation{Institute for Quantum Science and Technology, College of Science, National University of Defense Technology, Changsha Hunan 410073, China}
 \affiliation{Interdisciplinary Center for Quantum Information, National University of Defense Technology, Changsha Hunan 410073, China} 
 \affiliation{Hunan Key Laboratory of Mechanism and Technology of Quantum Information, Changsha Hunan 410073, China} 
  \author{Pingxing Chen}%
   \affiliation{Institute for Quantum Science and Technology, College of Science, National University of Defense Technology, Changsha Hunan 410073, China}
  \affiliation{Interdisciplinary Center for Quantum Information, National University of Defense Technology, Changsha Hunan 410073, China} 
  \affiliation{Hunan Key Laboratory of Mechanism and Technology of Quantum Information, Changsha Hunan 410073, China} 
  \author{Wei Wu}%
  \email{weiwu@nudt.edu.cn}
   \affiliation{Institute for Quantum Science and Technology, College of Science, National University of Defense Technology, Changsha Hunan 410073, China}
  \affiliation{Interdisciplinary Center for Quantum Information, National University of Defense Technology, Changsha Hunan 410073, China} 
  \affiliation{Hunan Key Laboratory of Mechanism and Technology of Quantum Information, Changsha Hunan 410073, China} 

\date{\today}

\begin{abstract}
In trapped ion system, accurate thermometry of ion is crucial for evaluating the system state and precisely performing quantum operations. However, when the motional state of a single ion is far away from the ground state, the spatial dimension of the phonon state sharply increases, making it difficult to realize accurate and mode-resolved thermometry with existing methods. In this work, we apply deep learning for the first time to the thermometry of trapped ion, providing an efficient and mode-resolved method for accurately estimating large mean phonon numbers. Our trained neural network model can be directly applied to other experimental setups without retraining or post-processing, as long as the related parameters are covered by the model's effective range, and it can also be conveniently extended to other parameter ranges. We have conducted experimental verification based on our surface trap, of which the result has shown the accuracy and efficiency of the method for thermometry of single ion under large mean phonon number, and its mode resolution characteristic can make it better applied to the characterization of system parameters, such as evaluating cooling effectiveness, analyzing surface trap noise.
\end{abstract}

\maketitle


\section{\label{sec:level1}Introduction}
Trapped ions have emerged as one of the most promising platforms for quantum computing, primarily attributed to their exceptional coherence time\cite{10.1063/1.5088164} and the ability to precisely control both internal and external states\cite{PhysRevLett.117.060504}. To thoroughly assess the performance of ion traps and facilitate coherent operations, it is crucial to accurately measure and control the effective temperature of ions.

For single trapped ion near the motional ground state, there exist several well-established techniques for thermometry, including sideband ratio method\cite{PhysRevLett.62.403}, singular value decomposition (SVD)\cite{PhysRevLett.76.1796}, etc. But these methods perform poorly when the motional state is far from the ground state. While in scenarios such as heating rate measurement\cite{PhysRevA.99.023412, PhysRevA.76.033411} and dissipative state preparation\cite{PhysRevLett.77.4728, kienzler2015quantum, lin2013dissipative}, thermometry is still necessary within a broader range of mean phonon number $\bar{n}$. To address this, alternative approaches, e.g., dark resonances\cite{Roßnagel_2015, PhysRevA.99.023412}, spatial thermometry\cite{PhysRevA.85.023427}, and Doppler recooling\cite{PhysRevA.76.033411} had been proposed. Dark resonances and spatial thermometry offer a measurement range from several hundred to ten thousand phonons, while Doppler recooling is specifically designed for $\bar{n} \ge 10^{4}$. Nevertheless, these techniques suffer from measurement complexity or inaccuracy\cite{RevModPhys.87.1419, PhysRevA.99.023412}. Furthermore, in many situations, it is necessary to separately measure the mean phonon number of certain vibration modes, while the aforementioned methods can only provide the estimation of the ensemble effective temperature composed by all relevant modes, instead of reliable information of single vibration mode. 

\begin{figure}[htpb]
\includegraphics[width=\columnwidth]{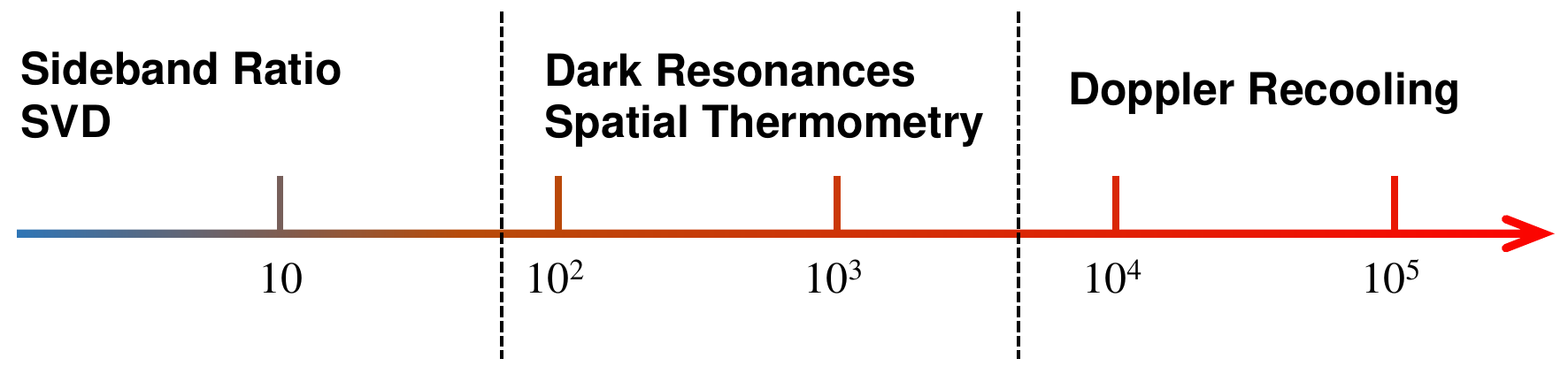}%
\caption{\label{fig:1}Application range of certain thermometry methods. }
\end{figure}

To tackle the challenges, we employ deep neural networks(DNNs) for the first time to estimate the mean phonon number of single vibration mode in single ion. Deep learning\cite{lecun2015deep} excels in analyzing intricate structures of high-dimensional data and extracting meaningful features from it. In the filed of natural sciences, deep learning has found diverse applications involving processing large amounts of complex data, such as identifying potential drug molecules\cite{doi:10.1021/acs.accounts.0c00699}, analyzing biological data\cite{10.1093/bioinformatics/btx531}, and its application in physics\cite{Sadowski2018, RevModPhys.91.045002} has mainly focused on particle physics\cite{ doi:10.1146/annurev-nucl-101917-021019}, quantum many-body physics\cite{PhysRevLett.122.250501}, and cosmology\cite{Firth2002EstimatingPR}.

In this work, we present an accurate, efficient, and mode-resolved method based on DNNs for the thermometry of single ion's motional state of $\bar{n}$ up to 1500 under thermal distribution, with no need for precise value of coupling strength $\Omega$. The data to be measured only involves excited state populations of multiple orders of blue sidebands\cite{wineland1998experimental,PhysRev.177.1857,PhysRevA.20.1521}, determined by the Lamb-Dicke parameter $\eta$, mean phonon number $\bar{n}$, coupling strength $\Omega$ and pulse duration $t$. By excluding the carrier transition which involves joint coupling with all existing vibration modes, it can avoid the effect from other vibration modes, thus realizing mode-resolved thermometry. 

The effective regime of our method spans from tens to over a thousand phonons, with the Lamb-Dicke parameter between 0.07 and 0.21, allowing for an adjustable measurement range of vibration frequency. We generated test datasets for theoretical test, which yielded an average error of less than 1\% on the noise-free test set and still below 10\% after introducing binomially distributed projection noise. Furthermore, we performed experimental verification of the method, where the error of the best model using 15 blue sidebands was found to be less than 10\%. It is noteworthy that the experimental operation of our thermometry method is remarkably straightforward, with results promptly obtained through the utilization of deep neural networks, eliminating the need for excessive post-processing.

\section{Method}
\subsection{Theory and training set}
 For a single ion trapped in a harmonic potential, we only consider its two internal states and axial harmonic vibration with frequency $\omega_{z}$, thus the state of the ion can be represented as $\left|\downarrow,m\right\rangle$ or $\left|\uparrow,n\right\rangle$ for phonon Fock state. The Rabi frequency $\Omega_{m,n}$ of transition  between these two states is given by \cite{PhysRev.177.1857,PhysRevA.20.1521}:
\begin{eqnarray}
\Omega_{m,n}&=&\Omega\left| \left\langle m\left| e^{i\eta\left(a+a^{\dagger}\right)}\right| n\right\rangle \right| \nonumber \\ 
 &=&\Omega e^{-i\frac{\eta^{2}}{2}} \eta^{m-n}\sqrt{\frac{n!}{m!}} L_{n}^{m-n}(\eta^{2}), \label{eq:1}
\end{eqnarray}
where we suppose $m \ge n$, and $\Omega$ is the coupling strength, $L_{n}^{\alpha}$ is the Laguerre polynomial, $\eta$ is the Lamb-Dicke parameter described by $\eta=\cos{(\theta)}k\sqrt{\hbar/2m\omega_{z}}$, which relates to ion mass $m$, wave number $k$, vibration frequency $\omega_{z}$, and angle $\theta$ between wave vector and vibration direction. 

Assuming that the ion is initially prepared in state $\left|\downarrow,n\right\rangle$, then weakly coupled to a blue-detuned laser, i.e. $\Omega\ll\omega_{z}$, the population of $\left|\uparrow,m\right\rangle$ state would be $P_{\left|\uparrow,m\right\rangle}=\sin^{2}(\Omega_{m,n}t)e^{-\gamma_{n} t}$ after an evolution duration $t$. Ulteriorly, when the motional state of the ion is a state described by a density matrix with diagonal elements being $p_{n}$, where $p_{n}$ represents the probability of $\left|n\right\rangle$, the excited state population of the $q$-th blue sideband will be
\begin{eqnarray}
P_{\uparrow}(q)=\sum_{n}p_{n}\sin^{2}(\Omega_{n+q,n}t)e^{-\gamma_{n} t},\label{eq:2}
\end{eqnarray}
in which the relaxation effect term $e^{-\gamma_{n}t}$ could be omitted, as the pulse duration $t$ in our method is much shorter than energy-level lifetime. Then $P_{\uparrow}(q)$ is rewritten as
\begin{eqnarray}
	P_{\uparrow}(q)=\sum_{n}p_{n}\sin^{2}(k_{n+q,n}\Omega t),\label{eq:3}
\end{eqnarray}
where $\Omega k_{n+q,n} = \Omega_{n+q,n}$ . With the knowledge of $\eta$, $p_{n}$, and $\Omega t$, the excited state probabilities of each blue sideband can be obtained using Eqs.~\eqref{eq:1} and \eqref{eq:3} straightforwardly. However, when the mean phonon number is relatively large, it is challenging to determine the accurate value of $\Omega$ experimentally by fitting the Rabi oscillating curve, due to the exceedingly small differences between $\Omega_{n+q,n}$ and $\Omega_{n'+q,n'}$ as $\bar{n}$ increases, thus making it difficult to get the exact value of $\bar{n}$. Meanwhile, in Eqs.~\eqref{eq:3}, the upper limit of $n$ needs to be high enough, in order to obtain reliable accuracy, e.g. for a thermal state whose $\bar{n} = 1500$, the upper limit of $n$ should reach 6000 to cover 98\% of the whole distribution, resulting in a quite time-consuming fitting process. 

\begin{figure}[htt]
	\centering
	\includegraphics[width=0.95\columnwidth]{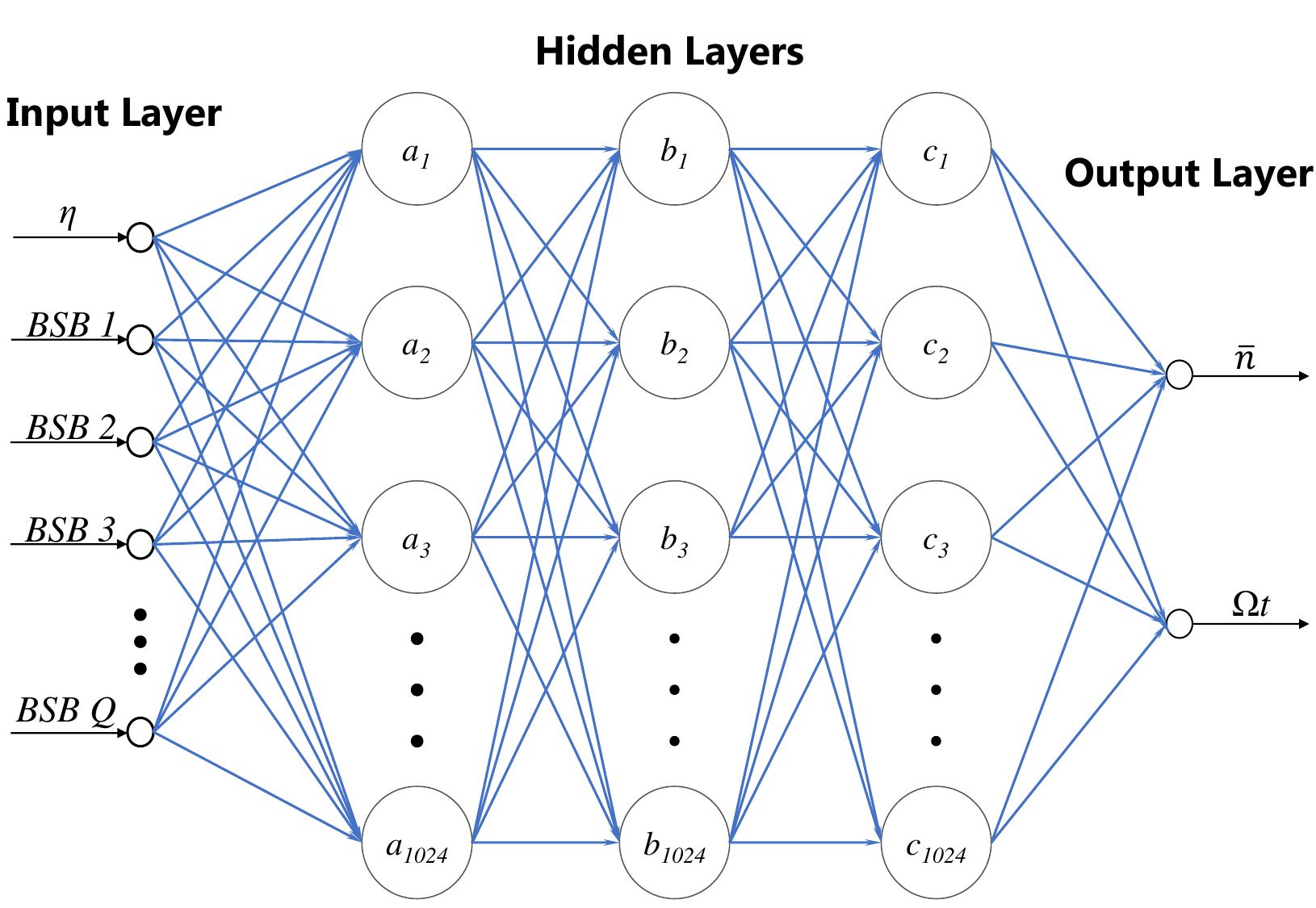}
	\caption{\label{fig:2}DNN structure of our work. The input layer involves of $\eta$ and  first to $Q$-th order of blue sideband, then there are three hidden layers each with 1024 neurons, $\bar{n}$ and $\Omega t$ are two outputs. }
\end{figure}

The essence of the above problem lies in the high dimension of phonon state of large $\bar{n}$. Thanks to the capability of DNNs to process complex data, this situation lends itself to a suitable application scenario of deep learning method. The initial step in constructing a DNN model involves generating a reliable training set. Typically, the dataset is derived from various sources such as public databases, experimental data, or theoretical calculation. However, in this application, due to the requirement of substantial amount of data for training and the necessity for precision, basically theoretical calculation is realistically feasible. Here, we narrow our focus to the thermal phonon state described by the probability distribution $p(m,\bar{n})=\bar{n}^{m}/(\bar{n}+1)^{m+1}$, for it is a widely concerned state. Considering our experimental configuration, we carefully choose discrete integer values ranging from 1 to 1500 as the variable $\bar{n}$. Furthermore, we impose restrictions on $\eta$, confining it to the range of 0.069 to 0.217, specifically tailored for the $^{40}$Ca$^{+}$ ion within our system. This choice stems from the fact that $\omega_{z}$ falls within the range of 200 to 2000 kHz when the laser's wave vector aligns parallel to the direction of vibration. Additionally, we have empirically determined that the interval of $0.5\pi$ to $2\pi$ for $\Omega t$ are experimentally suitable. Under various combinations of these parameters, we calculated $P_{\uparrow}(1)$ to $P_{\uparrow}(15)$. Each training data point comprises input values $\eta$,  $P_{\uparrow}(1), P_{\uparrow}(2), ..., P_{\uparrow}(Q)$, and outputs$\bar{n}$ and $\Omega t$. In order to ensure successful convergence during the training process, we have generated a dataset containing one million evenly-distributed samples, encompassing diverse parameter combinations.

In practical experiments, the electronic shelving technology allows for the identification of either a bright or dark state in each independent measurement. As a result, the measured probability $P_{k,N}$ follows a binomial distribution: 
\begin{eqnarray} 
P_{k,N}=\frac{k}{N}=C_{N}^{k}P_{\uparrow}^{k}(1-P_{\uparrow})^{N-k},\label{eq:4}
\end{eqnarray}
where $N$ represents the total number of measurements conducted while $k$ refers to the count of instances where an excited state is detected. Smaller $N$ brings greater noise, and it is essential for the model to exhibit robustness against such noise when dealing with limited $N$. 

Fig.~\ref{fig:2} illustrates the architecture of the DNNs employed in our study. By training with data involving different numbers of blue sideband, we have developed four distinct models, namely M-5, M-8, M-10, and M-15, respectively involving sideband numbers $Q=$ 5, 8, 10, and 15. Furthermore, to account for projection noise, we have trained an additional model called ANM-5, which is based on M-5 and retrained by a noisy training set. In this noisy training set, each probability $P_{\uparrow}$ is transformed into $P_{k,N}$ through a simulated binomial distribution process. This process utilizes $P_{\uparrow}$ along with a user-defined parameter $N$ to derive the count $k$. It is worth noting that all of these models follow the same training strategy and undergo an equal number of training epochs.

\subsection{Theoretical Performance}
 In order to preliminary verify these models, we randomly generate 50000 pieces of data as test set, with $\bar{n}$ ranging from 1 to 1500 and being decimal which are not included in the training set.

\begin{figure}[hb]
	\centering
	\begin{minipage}{0.5\linewidth}
		\includegraphics[width=\columnwidth]{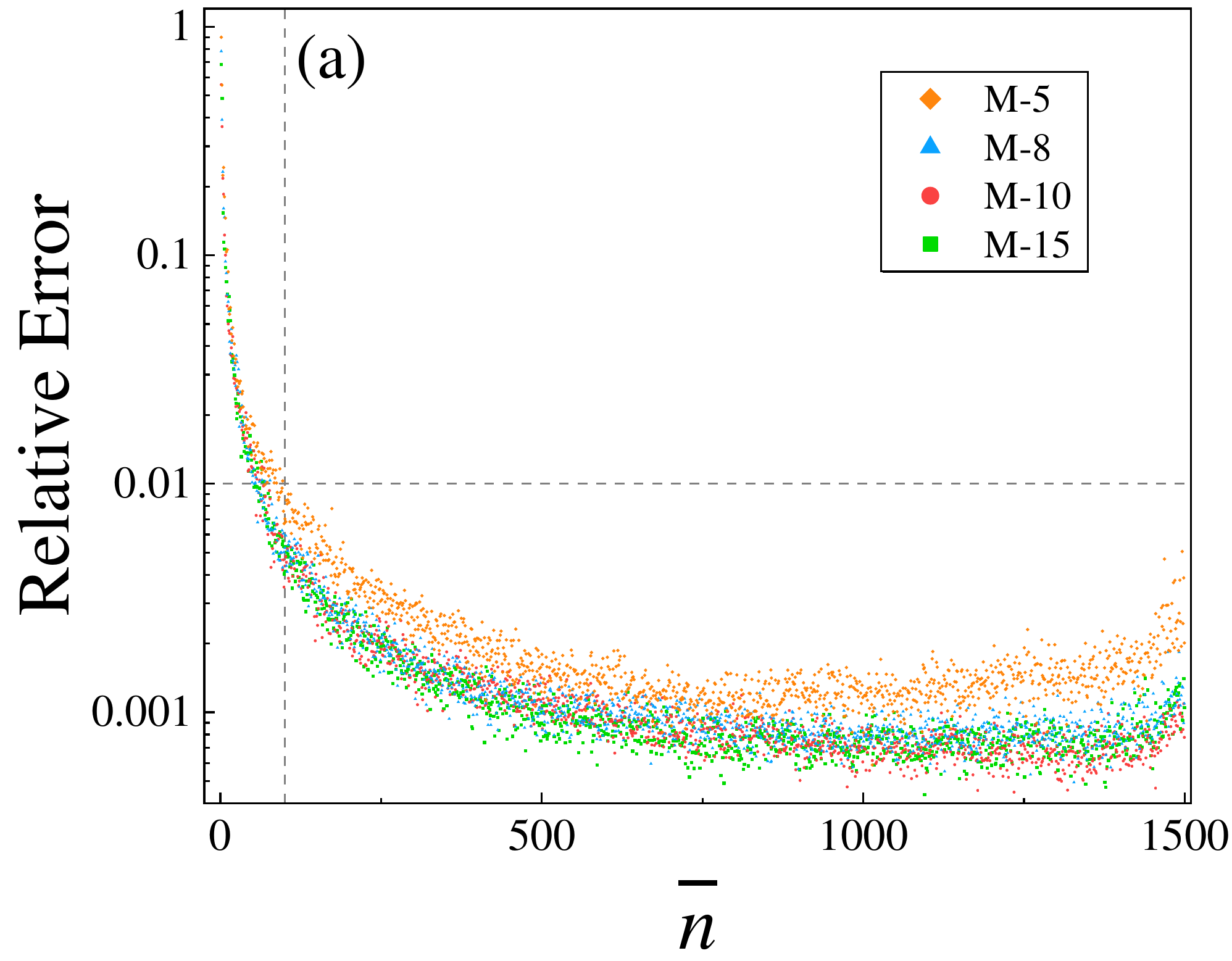}
		
	\end{minipage}
	\centering
	\begin{minipage}{0.485\linewidth}
		\includegraphics[width=\linewidth]{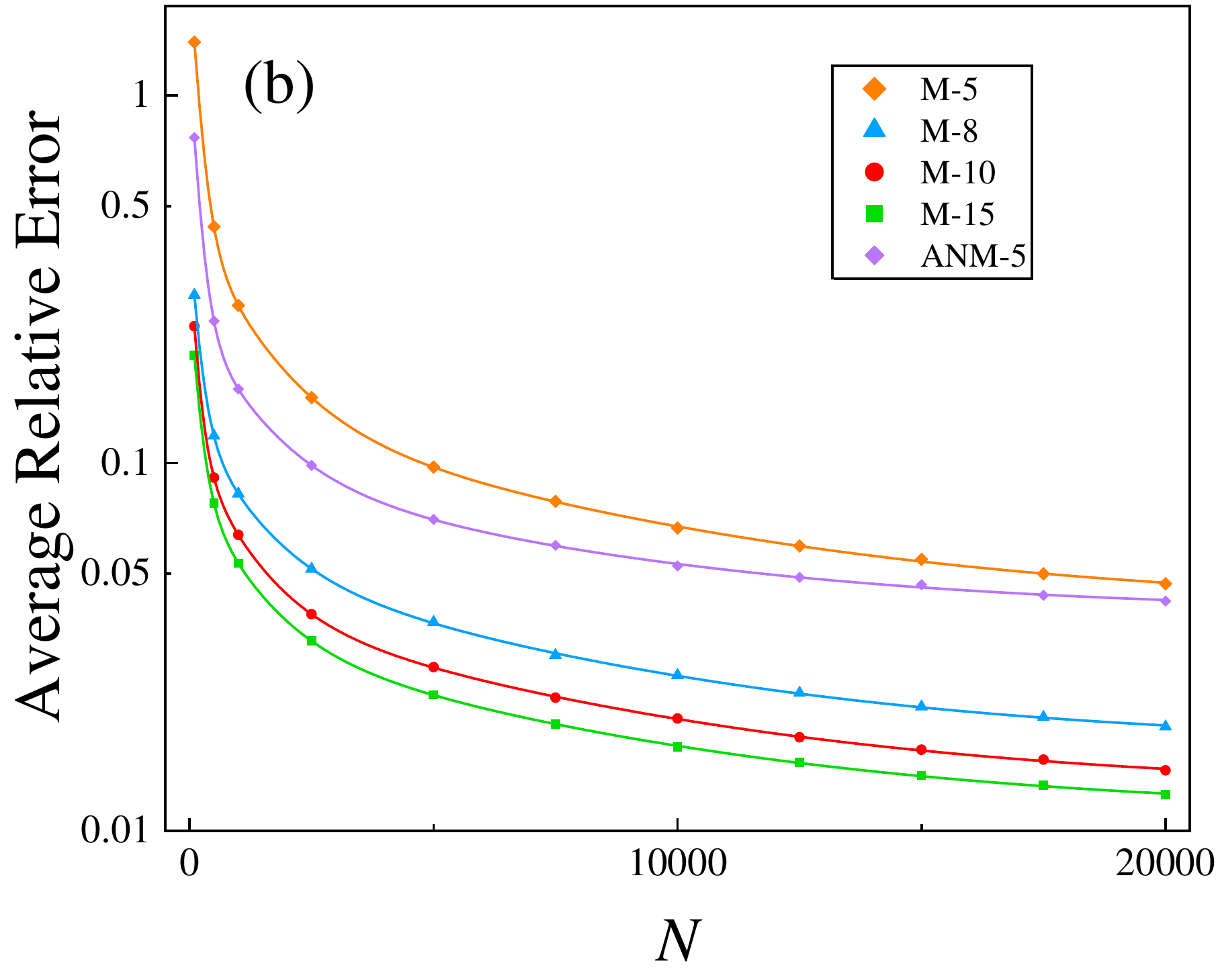}
		
	\end{minipage}
	\caption{\label{fig:3}Theoretical test results. (a)Relative error under different mean phonon numbers. When $\bar{n}$ is above 100, all models have an error of less than 1\%, while increased error is observed near the dataset boundary. (b)Average relative error on the entire test set with different total numbers of measurements $N$ in Eqs.~\eqref{eq:4}. ANM-5 is trained based on M-5 using training data with binomial noise. }
\end{figure}

\begin{figure*}[ht]
	\centering
	\begin{minipage}{0.49\linewidth}
		\includegraphics[width=\columnwidth]{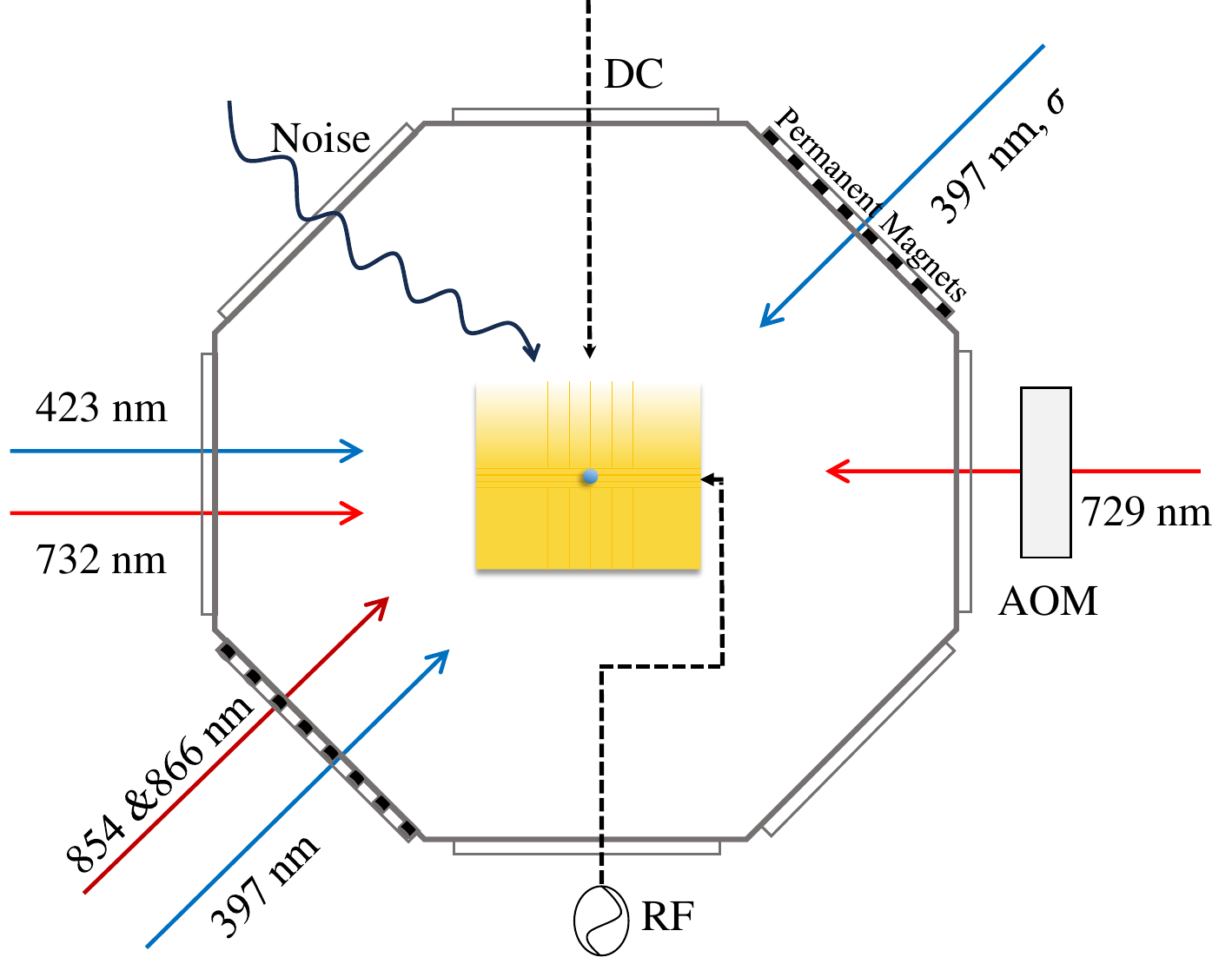}
		\centerline{(a)} 
	\end{minipage}
	\centering
	\begin{minipage}{0.5\linewidth}
		\includegraphics[width=\linewidth]{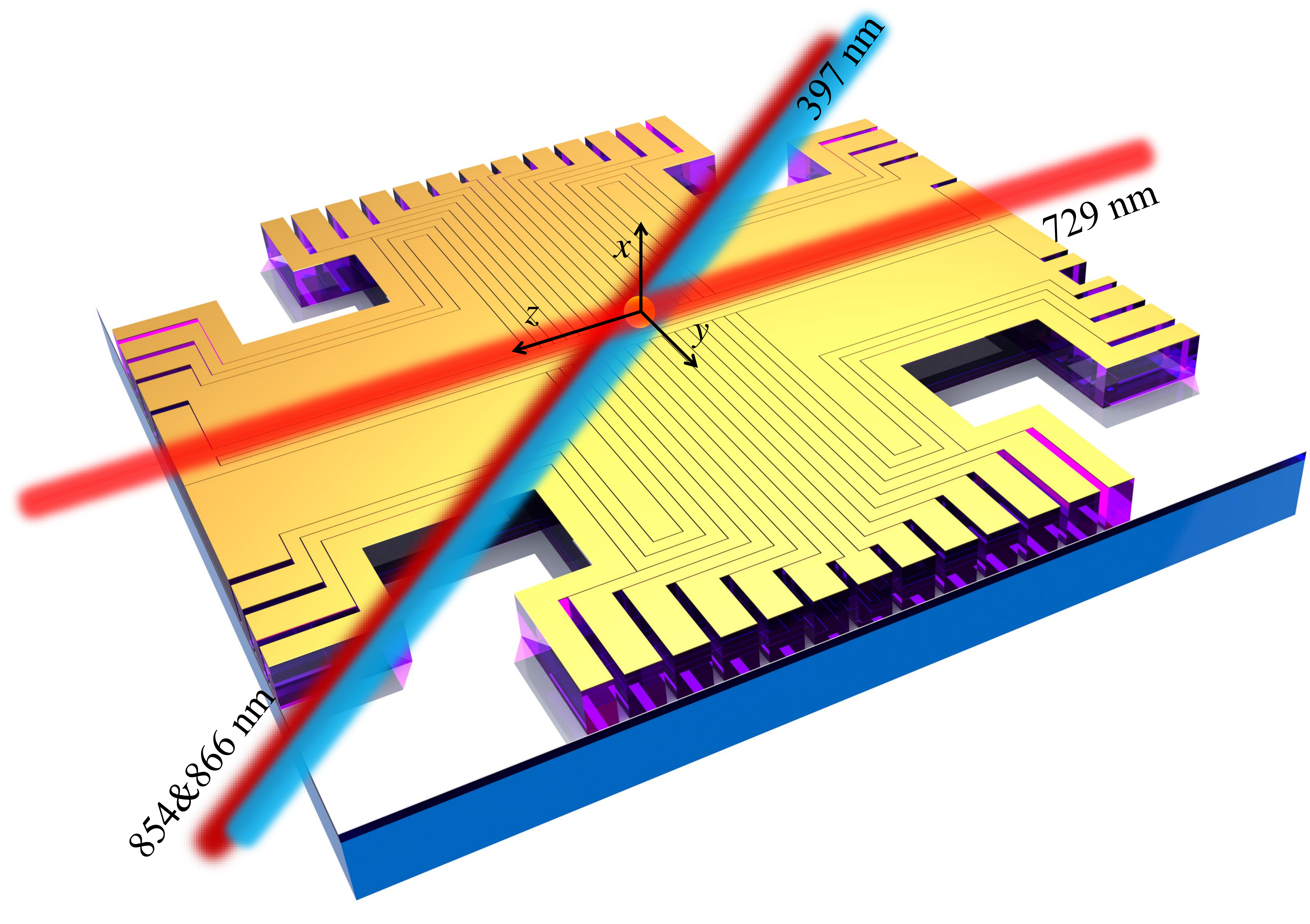}
		\centerline{(b)}
	\end{minipage}
	\caption{\label{fig:4}Experimental setup. (a)Optical and electrical settings.The dashed lines represent DC and RF voltages loaded on electrodes of the trap, and the black curve represents artificially added noise (b)Detailed model of surface trap. }
\end{figure*}
We first evaluated the performance of these DNN models under different mean phonon numbers, as illustrated in Fig.~\ref{fig:3}, where the relative error was represented by $\left|\Delta \bar{n}\right|/ \bar{n}$. Overall, for cases when $\bar{n}$ is below 8, all models exhibit a relatively big error ($\ge$ 10\%). This is primarily attributed to the fact that the excited state populations of some higher orders of sidebands approach zero within this $\bar{n}$ range, resulting in a lack of sufficient effective information. On the other hand, when the phonon number exceeds 100, the relative error of all models diminishes to less than 1\%, with a slight increase observed when $\bar{n}$ is approaching 1500. This increase is intrinsic to the limitations of deep learning, as data volume for training near the dataset boundary is relatively limited. Regarding the effects of utilizing different numbers of blue sidebands, M-15 and M-10 yield nearly identical results, while M-8 exhibits slightly inferior performance. Notably, there is a distinct increase in error for M-5, which may stem from the reduced information of higher orders of sidebands, making it challenging to obtain a smaller relative error.

Taking into account the inherent noise of the binomial distribution, we introduce different levels of such noise into the test set by varying total numbers of measurements $N$. As illustrated in Fig.~\ref{fig:3}(b), each data point represents the average relative error calculated across the entire test set. It's intuitive that the errors of all models decrease as noise decreases. Simultaneously, the error diminishes with increasing number of sideband involved, suggesting that employing more sidebands leads to enhanced robustness to the noise. Notably, ANM-5 which was trained using a noisy dataset shows improved performance. The fitting curve reveals that as $N$ increases, the error initially decreases rapidly before reaching a point of diminishing returns. As $N$ approaches infinity, the performance of ANM-5 does not significantly surpass that of the noise-free trained model, M-5, indicating that this training strategy still falls behind the gains achieved by increasing the number of sidebands.

The aforementioned analysis indicates that the method exhibits its effectiveness within the range of $\bar{n}$ from 100 to 1500, with its accuracy primarily contingent on experimental measurement precision, since the noise-free test shows good result. To achieve a relative error less than 5\%, it would require thousands of measurements to obtain a single value of population. However, to avoid measuring too many times in our practical experiment, this measurement method was replaced by peak fitting, as further elaborated in the subsequent sections.

\section{Experimental verification}
\subsection{Experimental scheme and setup}
The experiment was conducted on a six-wire surface trap, which can flexibly adjust trap frequency and efficiently prepare target thermal states for verification, as will be further explained below. This trap has 15 pairs of DC electrodes, of which 7 pairs are mainly used to trap ions. The single $^{40}$Ca$^{+}$ ion is trapped at a distance of approximately 150 $\mu$m above the chip surface with the axial frequency $\omega_{z}$ set to 626 kHz, by applying a radio frequency voltage of 20.9 MHz and DC voltages. Other DC electrodes are given appropriate voltages to compensate for micro-motion.

Two Zeeman sublevels of $^{40}$Ca$^{+}$ ion, $^{2}S_{1/2}(m_{j}=-1/2)$ and $^{2}D_{5/2}(m_{j}=-5/2)$, are used as ground $\left|\downarrow\right\rangle$ and excited states $\left|\uparrow\right\rangle$, which is generated by a magnetic field whose direction is at an angle of 45 degrees to the axial direction. The state manipulation is implemented by a 729 nm laser parallel to the axial direction, with its switch and frequency controlled by Acousto-Optic Modulator (AOM). With $\omega_{z}$ and laser direction, we determine a Lamb-Dicke parameter $\eta$ of approximately 0.122, which is adjustable by changing the DC voltages to change $\omega_{z}$. The cooling methods include Doppler cooling and sideband cooling. 397 nm and 866 nm lasers are used for Doppler cooling, and sideband cooling uses 729 nm and 854 nm lasers, by which the ion can be cooled to about 0.3 phonons. Under the above conditions, the measured heating rate is 0.245$\pm$0.046 ms$^{-1}$. Due to the relatively low heating rate of our surface trap, and for the sake of experimental convenience, we additionally introduced a 5 V$_{pp}$ noise signal to the 15th pair of electrodes to expedite the heating process.

\begin{figure}[h]
	\centering
	\includegraphics[width=\columnwidth]{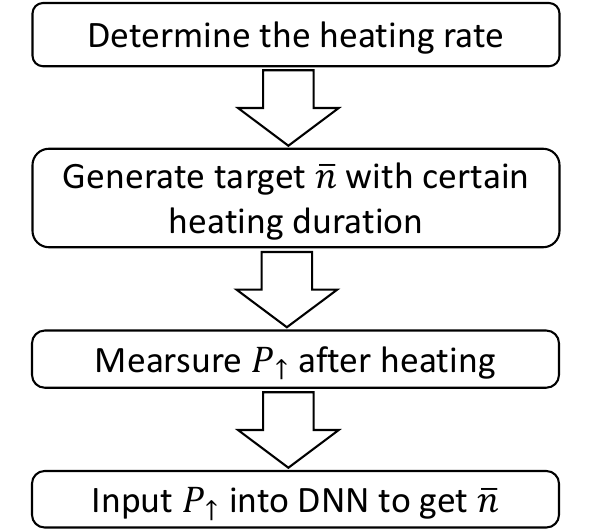}
	\caption{\label{fig:5}The experimental verification procedure}
\end{figure}
In order to ascertain the effectiveness of this approach, it is crucial to accurately generate thermal states of various mean phonon numbers. The current experimental situation provides a good initial state, and then combined with a constant heating rate, we can accurately prepare thermal states of any $\bar{n}$ we need. The experimental verification procedure unfolded as follows: 

\begin{figure}[h]
	\centering
	\includegraphics[width=\columnwidth]{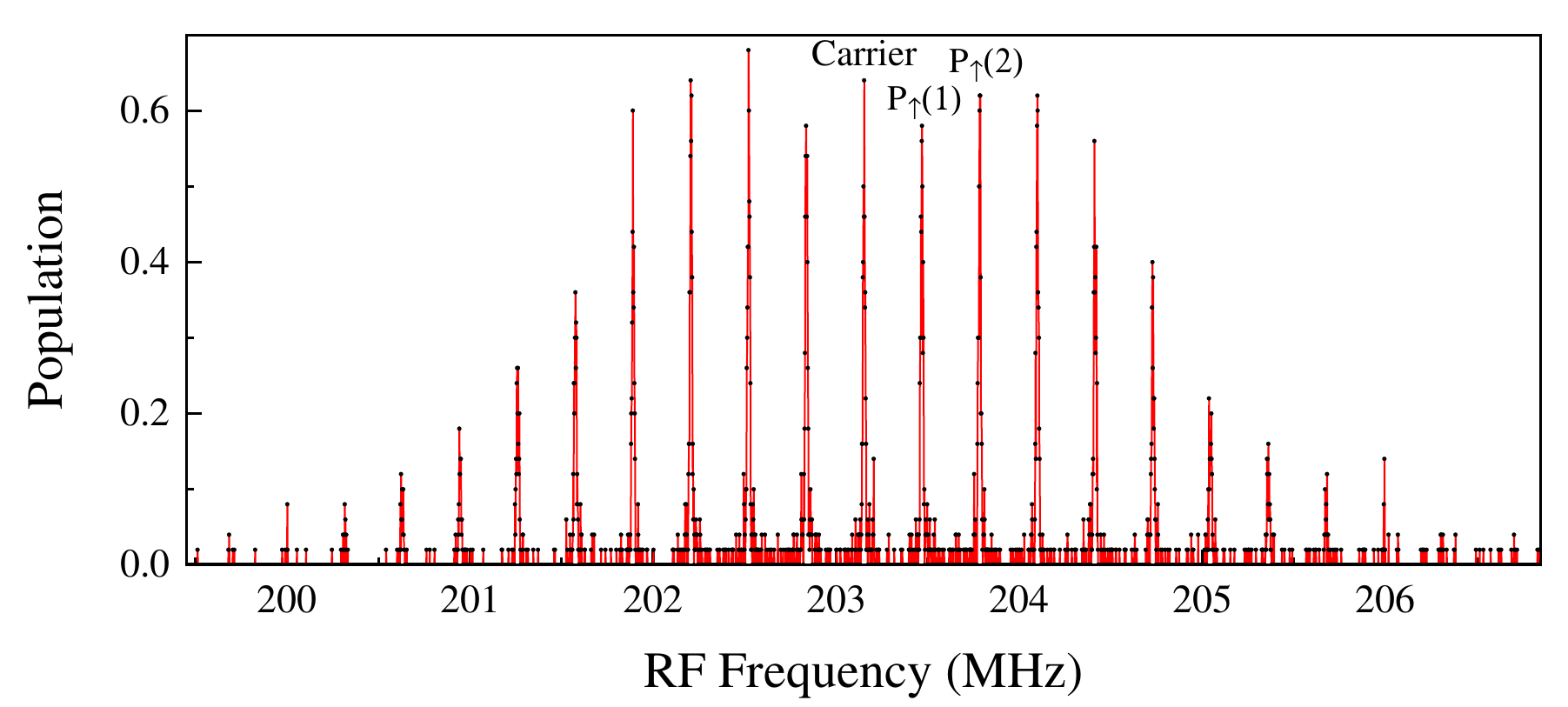}
	\caption{\label{fig:6}Excitation spectrum of motional sidebands. RF signal is loaded on AOM to control the laser frequency. In practical experiment, there's no need to scan such a large range, but finely scan small areas around each sideband and find the peaks with Gaussian fitting, by which frequency error and AC-Stark shift can be ruled out.}
\end{figure}

 1. We measured $\bar{n}$ using the Rabi oscillation fitting method under various heating duration near the motional ground state, for the fitting method was deemed accurate within the specified range, allowing us to obtain the accurate heating rate, and we considered that this manufactured heating rate remained constant over time.
 
 2. Based on the determined heating rate, subject the ion to an extended heating duration to attain the target thermal state with a higher mean phonon number, by which we can have an accurate value of $\bar{n}$ for reference.
 
 3. Rightly after the heating, we controlled the frequency of 729 nm laser to measure $P_{\uparrow}$ up to 15th blue sideband, as shown in Fig.~\ref{fig:5}. It is worth noting that, to minimize measurement error, we conducted a small-frequency-scale scan around each resonance point, followed by peak fitting using Gaussian distribution to determine the final value of population.
 
 4. Upon completion of the experiment, input the measured population values into the DNN model and compare the model's output with the reference value.

Due to the artificially introduced constant heating rate, this experimental scheme ensures the precision for generating thermal states with different mean phonon numbers and provides a reliable verification of our method.

\subsection{Result}
\begin{figure}[h]
	\centering
	\includegraphics[width=\columnwidth]{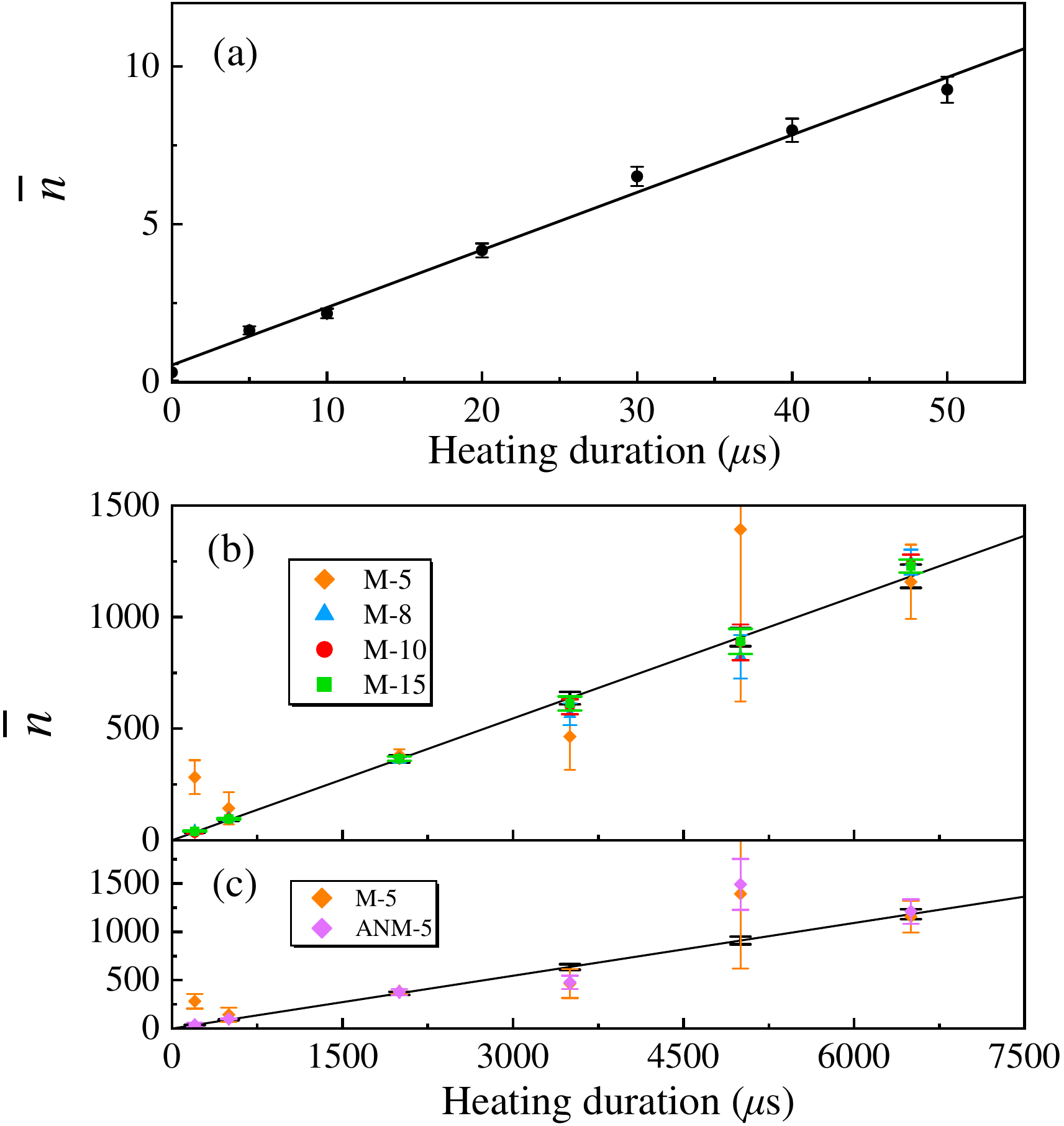}
	\caption{\label{fig:7}Experiment results. (a)Heating rate measured by fitting Rabi oscillation. (b)DNN result in large phonon number scale. (c) Comparison of M-5 and ANM-5. The black lines in (b) and (c) are the extrapolated heating line in (a).}
\end{figure}

The heating rate measured using the method of fitting Rabi oscillation is shown in Fig.~\ref{fig:7}(a). By adding noise signal, we increase the heating rate from 0.245$\pm$0.046 ms$^{-1}$ to 182$\pm$8 ms$^{-1}$. Based on this artificially introduced heating rate, we selected several heating duration points for verification. 

Under different heating duration, thermal states of different $\bar{n}$ were obtained, and then the populations of multiple orders of sidebands for different thermal states were measured as input to DNN models. The output results generated by the neural networks are illustrated in Fig.~\ref{fig:7}(b). Within the phonon number range spanning from tens to over a thousand, models employing 8 or more sidebands have comparable performance and demonstrate a commendable alignment with the extrapolated heating line. In contrast, M-5 exhibits relatively poor performance, largely attributed to its unresistant to the heightened noise levels encountered in real-world experiments. Yet, M-5 might yield improved results with more measurement times, given that its theoretical performance is acceptable in above theoretical noise-free test. 

The error bars in the figure were generated using Monte Carlo simulation\cite{4736059}, where we employ the uncertainties introduced during peak fitting as standard deviations in a normal distribution. Then a series of input values generated by simulating such normal distribution were fed into the DNN model to obtain a set of outputs whose standard deviation was taken as the error bar. Therefore, the size of error bars is influenced by both the accuracy of measurements and the inherent characteristics of the neural networks. Remarkably, an increase in the number of sidebands employed leads to a reduction in the size of the error bars, signifying that a greater number of sidebands contributes to enhanced robustness, which aligns with the theoretical test presented in previous section. The above analysis suggests that using more sidebands brings better performance, but since the difference among M-8, M-10 and M-15 is not significant, it's feasible to use M-10 for conveniency in practical experiments.

Furthermore, ANM-5 was subjected to testing, shown in Fig.~\ref{fig:7}(c). Overall, training under projection noise resulted in a reduction in error bar. In low $\bar{n}$ range, its performance surpassed that of M-5, although there was minimal discernible difference in the higher range. The lack of substantial improvement, especially at the 5000 $\mu$s point, may be attributed to inherent defects in M-5 that persisted even after retraining with noisy data. 
 
\section{Summary}
In this work, we showcased the utilization of Deep Neural Networks for precise thermometry of single ion. Our method has demonstrated remarkable performance across a range of mean phonon number from tens to well over a thousand. Additionally, we explored the impact of employing different numbers of sidebands as inputs, revealing that increased sidebands enhance both accuracy and robustness. Model trained with noisy data exhibits certain improvements in robustness, but falls short of the gains achieved by increasing the number of sidebands. For practical applications balancing between efficiency and accuracy, adopting 10 sidebands is deemed sufficient within the above mentioned $\bar{n}$ range.

It's essential to note that this work only focused on the thermal distribution of single ion's phonon state. But in principle, the method's applicability extends to various other distributions. When confronted with different ranges and distributions, creating tailored datasets for training new neural networks suffices. Furthermore, this method is highly adaptable across different experimental setups, requiring no modifications to the opto-electronic settings and independent of the ion type, which facilitates swift deployment. Benefited by the efficiency inherent in neural networks, results can be rapidly figured out following the acquisition of experimental data.

In summary, we present an accurate, efficient, and mode-resolved approach for estimating the mean phonon number of single trapped ion, particularly in case of large $\bar{n}$ range. In the future, we envision extending this method to accommodate scenarios such as estimation of phonon coherent state, feedback control of system setups, and more crucially, the thermometry of multiple ions.

\begin{acknowledgments}
This work is supported by the National Natural Science Foundation of China under Grants No.12174447, No.11904402, No.12004430, No.12074433, No.12174448, and No.12204543, and the Science and Technology Innovation Program of Hunan Province under Grant No.2022RC1194.
\end{acknowledgments}

\appendix*
\section{Design of DNNs}
A typical DNN model usually consists of an input layer,some hidden layers, and an output layer. The dimensions of the output and output layers are determined by actual requirements, and parameters such as the number of hidden layers and the number of neurons in each layer are manually tuned. The structural parameters of the network, training strategies as well as  the size and quality of the training set, all affect the effectiveness of the model. To build the DNNs, here we use PyTorch\cite{NEURIPS2019_bdbca288}, an open-source deep learning framework known for its excellent flexibility and ease of use.

The neural network in this work composed of an input layer, 3 hidden layers, and an output layer. Including the Lamb-Dicke parameter, the input values naturally range from 0 to 1, making it suitable as input for the network without any post-processing. There are 1024 neurons in each hidden layer, which is enough for successfully converging. The activation functions use Tanh and ReLU\cite{Glorot2011DeepSR}, which can introduce non-linearity. 

For the large order of magnitude difference between the two output values, which is not conducive to evaluating the convergence effect and updating weights, the range of their values are scaled to the same interval: [-10,10]. To evaluate the distance between the predicted value and the target value, we use the following loss function:
\begin{eqnarray}
	loss(x,y)=\frac{\left| y_{1}-f(x)\right| +\left| y_{2}-g(x)\right| }{2},\label{eq:5}
\end{eqnarray}
where $x$ represents input, $f(x)$ and $g(x)$ are outputs and 
\begin{eqnarray}
	y_{1}&=&\frac{\bar{n}}{75}-10,  \nonumber \\ 
	y_{2}&=&\frac{15}{2}(\frac{\Omega t}{\pi}-\frac{1}{2})-10. \label{eq:6}
\end{eqnarray}

By continuously iterating through loops, the specific parameters inside networks can be updated automatically by the optimizer, where Adam\cite{kingma2017adam} was adopted, until the loss converged, for example, the loss of M-15 converged to 0.009 after 20 epochs of training.

\bibliographystyle{unsrt}
\bibliography{notes}

\end{document}